\documentclass[11pt,letterpaper]{article}
\pdfoutput=1
\usepackage{jheppub}
\usepackage{color}
\usepackage{graphicx}
\usepackage{wrapfig}
\usepackage{amssymb,amsmath}

\usepackage[utf8]{inputenc}

\def\beq{\begin{equation}}
\def\eeq#1{\label{#1}\end{equation}}
\def\eeqn{\end{equation}}
\def\beqa{\begin{eqnarray}}
\def\eeqa#1{\label{#1}\end{eqnarray}}
\def\eeqan{\end{eqnarray}}

\def\leqn#1{(\ref{#1})}

\title{Infrared Safety of a Neural-Net Top Tagging Algorithm}

\author[a]{Suyong Choi,}
\author[a,b]{Seung J. Lee,}
\author[c]{Maxim Perelstein}

\emailAdd{suyong@korea.ac.kr}
\emailAdd{sjjlee@korea.ac.kr}
\emailAdd{mp325@cornell.edu}

\affiliation[a]{Department of Physics, Korea University, Seoul 02841, Republic of Korea}
\affiliation[b]{School of Physics, Korea Institute for Advanced Study, Seoul 130-722, Korea}
\affiliation[c]{Laboratory for Elementary Particle Physics, Cornell University, Ithaca, NY 14853, USA}

\abstract{Neural network-based algorithms provide a promising approach to jet classification problems, such as boosted top jet tagging. To date, NN-based top taggers demonstrated excellent performance in Monte Carlo studies. In this paper, we construct a top-jet tagger based on a Convolutional Neural Network (CNN), and apply it to parton-level boosted top samples, with and without an additional gluon in the final state. We show that the jet observable defined by the CNN obeys the canonical definition of infrared safety: it is unaffected by the presence of the extra gluon, as long as it is soft or collinear with one of the quarks. Our results indicate that the CNN tagger is robust with respect to possible mis-modeling of soft and collinear final-state radiation by Monte Carlo generators.     
}

\begin{document}

\maketitle	

\section{Introduction} 
\label{sec:introduction}
Events at the Large Hadron Collider (LHC) contain large numbers of jets. The jets can be classified into four types, according to their origin: (i) Light-quark jets, initiated by $u, d, s$ or $c$ quarks; (ii) Gluon jets; (iii) $b$-quark initiated jets; and (iv) jets created by a hadronic decay of a highly boosted massive object, such as a $W/Z$ boson, Higgs, or top quark. In the latter case, hadronic showers created by each of the partons overlap, and standard jet reconstruction algorithms recognize them as a single merged jet. A {\it jet classification algorithm}, or ``tagger", attempts to reconstruct the origin of each individual jet, based on the information accessible to the experiment, {\it i.e.} detector-level data. Recently, there has been strong interest in applying modern machine-learning techniques, such as Neural Networks (NNs), to the jet classification problem. This is motivated as follows. The pattern of energy deposits in individual hadron calorimeter (HCAL) cells can be thought of as a two-dimensional image of the jet. Jets of each type have a characteristic shower history, resulting in differences in spatial distribution of energy inside the jet, often called ``jet substructure". The jet classification problem is thus mapped onto a 2D image recognition problem~\cite{Cogan:2014oua}. Application of NNs to image recognition is a well-developed field of computer science. Advanced NN-based image recognition techniques have been applied to jet classification problems in Monte Carlo (MC) studies, with highly promising results~\cite{Almeida:2015jua,deOliveira:2015xxd,Baldi:2016fql,Barnard:2016qma,Komiske:2016rsd,Kasieczka:2017nvn}. (For related applications of machine-learning techniques, see Refs.~\cite{Louppe:2017ipp,Pearkes:2017hku,Butter:2017cot,Metodiev:2017vrx,Datta:2017lxt,Cheng:2017rdo,Egan:2017ojy,Luo:2017ncs,Macaluso:2018tck,Fraser:2018ieu,Roxlo:2018adx,Collins:2018epr}; for a recent review, see~\cite{Larkoski:2017jix}). For example, NN-based top taggers have been shown to significantly outperform traditional top-tagging algorithms currently in use by the LHC experiments.              

Can NN-based taggers trained on MC samples be used in real data analysis? The answer hinges on whether the features of jet substructure that are identified by the NN as important for classification are in fact accurately modeled by the MC generator. This is a non-trivial issue. Parton showering cannot be described by fixed-order perturbation theory, since soft and collinear parton splittings suffer from infrared/collinear (IRC) singularities. As a result, MC predictions of energy distribution within jets, in particular on small angular scales, suffer from significant (and poorly quantified) theoretical uncertainties. At the same time, unlike traditional taggers, the highly non-linear, multi-variable nature of the NN tagger output makes it very difficult to identify the specific features in the jet substructure that the NN focuses on, let alone assess their robustness in the simulation. To date, this issue has been addressed by cross-comparisons of NN taggers trained on samples produced by different MC generators, which employ different algorithms to model parton showers (see~{\it e.g.} Refs.~\cite{Almeida:2015jua,Barnard:2016qma}). While the results seem to indicate that the NN output is robust, a deeper understanding of this issue is clearly desirable to put this approach to jet tagging on a firm foundation.\footnote{An alternative would be to avoid the use of MC generators altogether by training directly on real data. This would require one to identify training sets, tagged by an object external to the jet, in the data. For recent interesting work in that direction, see Ref.~\cite{Metodiev:2017vrx,Komiske:2018oaa}.}      

\begin{figure}[t!]
	\begin{center}
		\includegraphics[width=0.35 \linewidth]{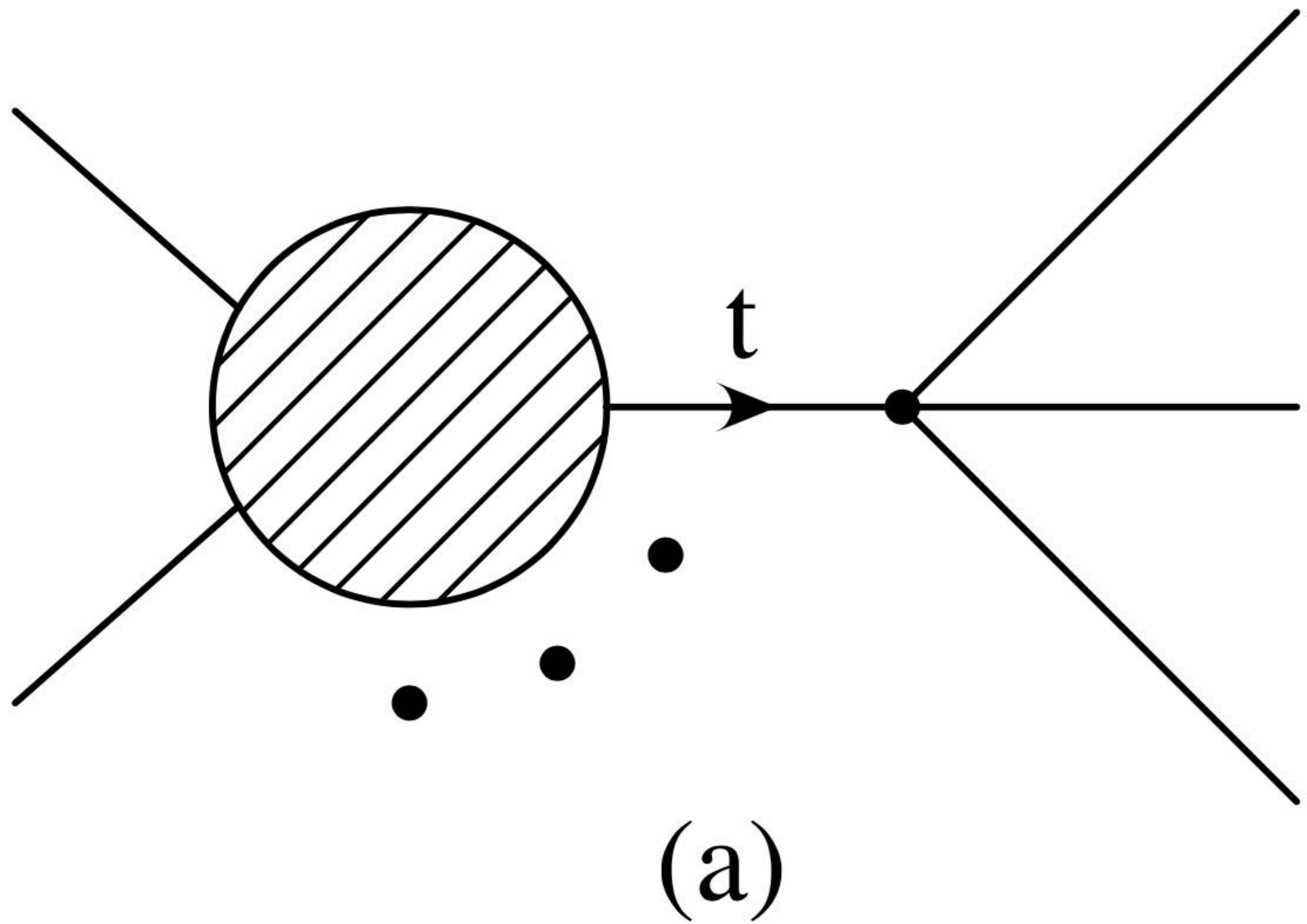}
   		\hspace{1cm}
   		\includegraphics[width=0.35 \linewidth]{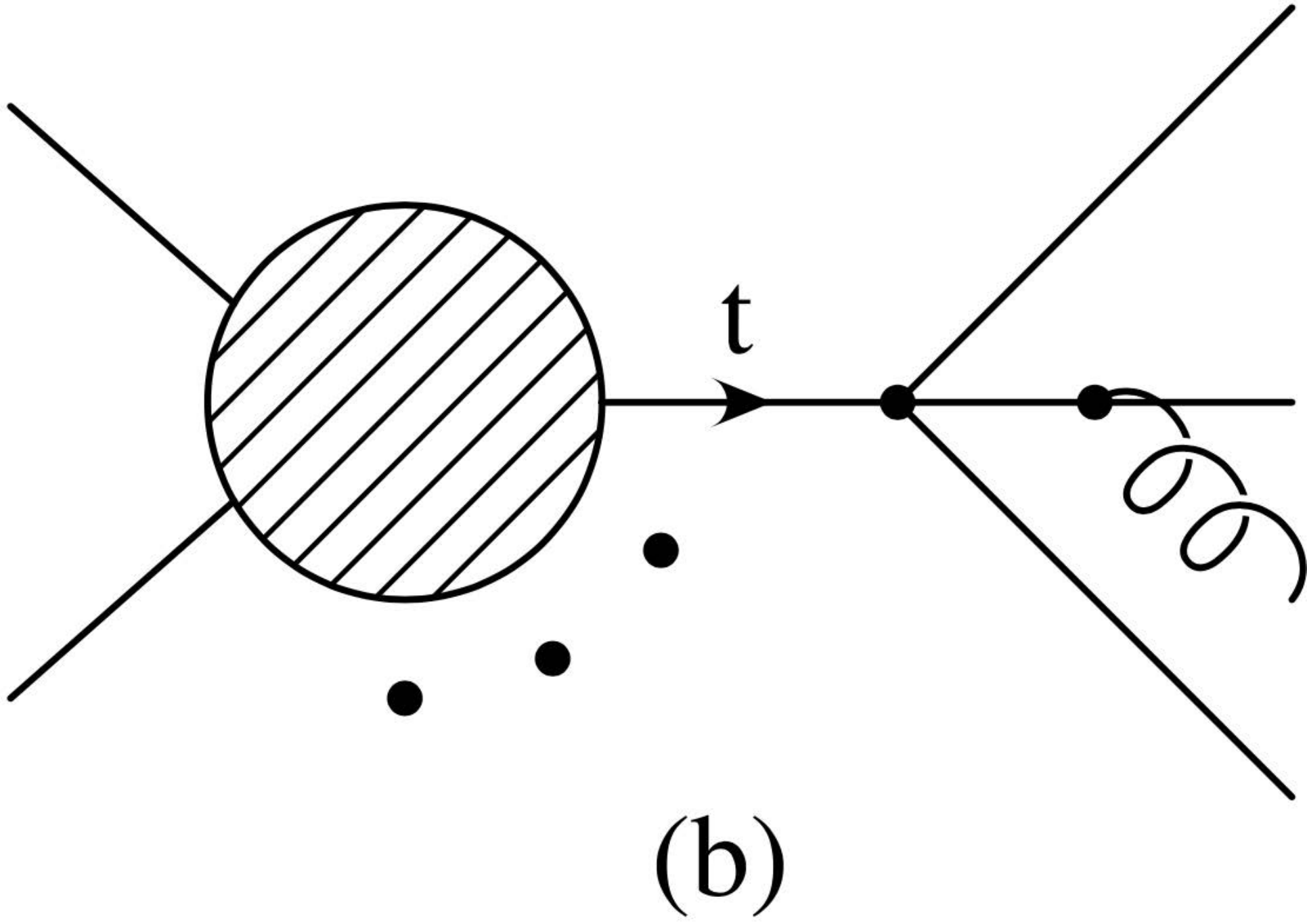}
	\end{center}
	\caption{\label{fig:TopGem} Parton-level top decays, with and without a final-state gluon respectively.}
\end{figure}

Traditionally, observables in jet physics are thought to be robust with respect to uncertainties in parton shower modeling if they satisfy the requirement of {\it Infrared/Collinear (IRC) Safety}. The notion of IRC safety applies to parton-level events. An observable ${\cal O}$ is IRC safe if a soft or collinear splitting of one of the partons leaves ${\cal O}$ unchanged:
\beq
{\cal O}_n(p_1,\ldots, p_i, p_{i+1},\ldots, p_n)\rightarrow{\cal O}_{n-1}(p_1,\ldots,p_i+p_{i+1},\ldots ,p_n)
\eeq{IRS_def}
whenever $p_{i+1}$ becomes soft or collinear with $p_i$. For example, consider the two events shown in Fig.~\ref{fig:TopGem}. In the limit when gluon in Fig.~\ref{fig:TopGem} (b) becomes either soft ($p_{T,g}\to 0$) or collinear with one of the quarks $(p_g\cdot p_i\to 0)$, the value of ${\cal O}$ evaluated on the final state (b) should approach its value evaluated on the final state (a).    

It's worth noting that IRC safety is not the necessary condition for calculability of physical observables. Sudakov safe observable~\cite{Larkoski:2013paa, Larkoski:2015lea} is a notable example which is IRC unsafe, but calculable if all-orders effects are resummed. The numerical analysis of this paper cannot distinguish between IRC-safe and ``unsafe but calculable", Sudakov-safe observables, because Sudakov-suppressed regions of phase space are not probed. While we will use the term ``IRC safety" throughout the paper for brevity, this caveat should be kept in mind.  

NN tagger is an observable that maps the matrix of energy deposits in individual HCAL cells onto a number between 0 and 1, the ``topness" of the jet. The goal of this paper is to check whether this observable is IRC safe. We perform this test in the particular context of a Convolutional Neural Network (CNN) top tagger. The CNN is first trained on {\it particle-level} (showered and hadronized) MC samples of boosted top jets and ``QCD" (light quark/gluon) jets. We then apply this CNN to {\it parton-level} hadronic top events. This defines a parton-level observable, to which the above canonical definition of IRC safety can be applied. We study the behavior of this observable as a function of the gluon momentum and collinearity in the $(t\to 3q)+g$ sample shown Fig.~\ref{fig:TopGem} (b). Our numerical results strongly support the hypothesis that {the CNN output is IRC-safe.} The training process appears to result in a network that largely disregards small-scale angular features in the energy distribution inside the jet, making the CNN tagger robust with respect to modeling such small-scale features in MC generators. Such robustness is a necessary pre-condition for practical applicability of MC-trained NN taggers, and it is highly reassuring that it is satisfied. 

The rest of the paper is organized as follows. In Section~\ref{sec:setup}, we discuss the architecture of the CNN tagger, its training and performance on particle-level MC samples. We also describe the parton-level ``merged" and ``unmerged" samples used for numerical tests of IRC safety of the tagger. The main results of the analysis are presented in Section~\ref{sec:results}, which contains the evidence to support our claim that the CNN observable is IRC-safe according to the canonical definition, Eq.~\leqn{IRS_def}. Discussion of the results and conclusions are contained in Section~\ref{sec:discussion}.

\section{Neural Net Tagger and Event Samples}
\label{sec:setup}
\begin{figure}[t!]
	\begin{center}
		\includegraphics[width=14cm]{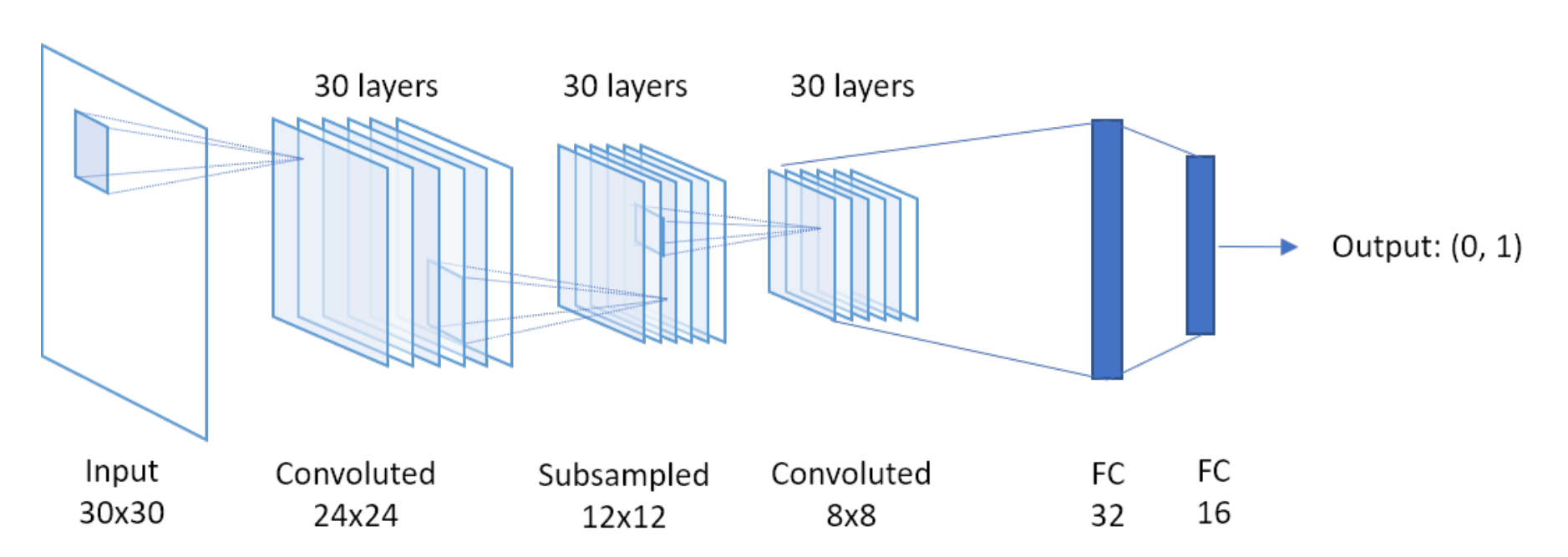}
	\end{center}
	\caption{\label{fig:cnnarch} Convolutional Neural Network (CNN) architecture used for boosted top-jet tagging.}
\end{figure}

The top tagger used in this study consists of a Convolutional Neural Network (CNN), which has proven to be one of the best performers in problems of pixelated image recognition. CNN architecture is known to produce robust identification of translationally invariant features of {\it a priori} unknown size. In our case, the feature of interest is subjets, which clearly play an important role in top tagging. The network architechture is schematically shown in Fig.~\ref{fig:cnnarch}. We used the {\tt mxnet} software package for implementation of the CNN~\cite{mxnet} on a NVIDIA Geforce GTX 1080 GPU. The input layer of the CNN is the Hadronic Calorimeter (HCAL), modeled as a set of $30\times 30$ square pixels of size $(\Delta\phi, \Delta\eta)=(0.1, 0.1)$. The pixels are populated by normalized energy deposited in each bin by a jet, preprocessed according to the procedure used in Ref.~\cite{Almeida:2015jua}. Preprocessing places the center of the jet at the center of the image, and rotates the jet so that the principal axis always has the same orientation. In this way, overall translational and rotational symmetries of the jet are factored out and do not need to be learned in the training process.\footnote{As an alternative, one can use $p_T$, instead of energy, as the input observable, motivated by Lorentz invariance~\cite{deOliveira:2015xxd}. We explored this alternative and did not find a significant difference in training time or performance.} The next layer consists of thirty ``filters", which are convoluted with the input image according to 
\begin{equation}
o_\ell (i,j)=\sum_{x=1}^7 \sum_{y=1}^7 I(i+x, j+y) w_\ell (x,y) \qquad (i,j=1,\ldots,24,\  \ell=1,\ldots,30),
\end{equation}
where $I$ is the input image, $w_\ell$ are the filters, and $o_\ell$ is the output of the convolution operation. We use {\tt RelU} activation function. The individual weights of the filters, $w_\ell (x,y)$ are determined during training of the CNN, using back-propagation methods. Each filter is to learn some distinguishing features that separate signal from backgrounds. The outputs of these layers are subsampled  and convoluted further with different set of filters, as shown in Fig.~\ref{fig:cnnarch}. Ultimately, the final fully connected layer produces a single output, the ``topness" of the jet $Y\in [0, 1]$, with $Y=1$ corresponding to a boosted top jet and $Y=0$ corresponding to a QCD jet. 

\begin{figure}[t!]
	\begin{center}
		\includegraphics[width=0.45 \linewidth]{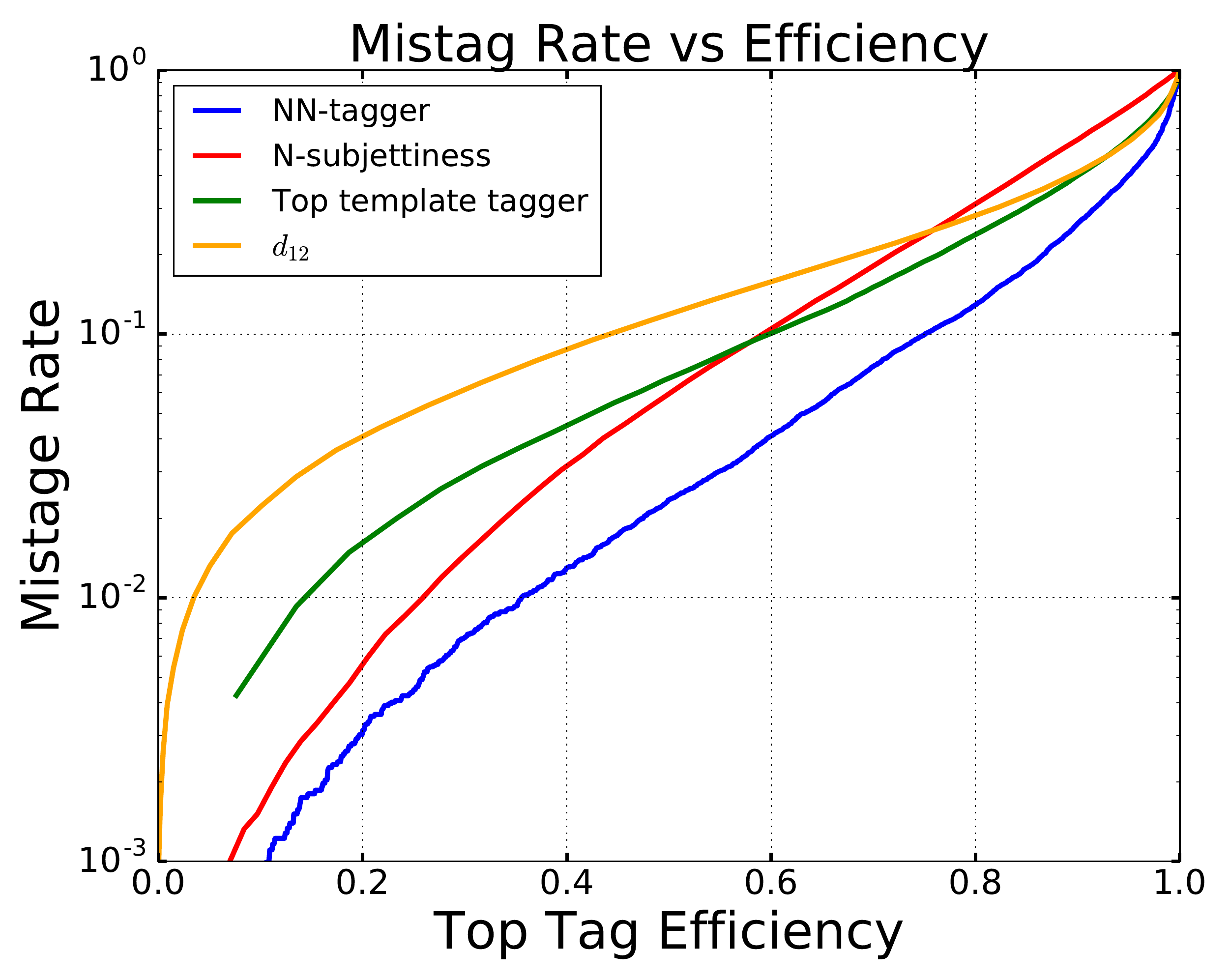}
		\includegraphics[width=0.45 \linewidth]{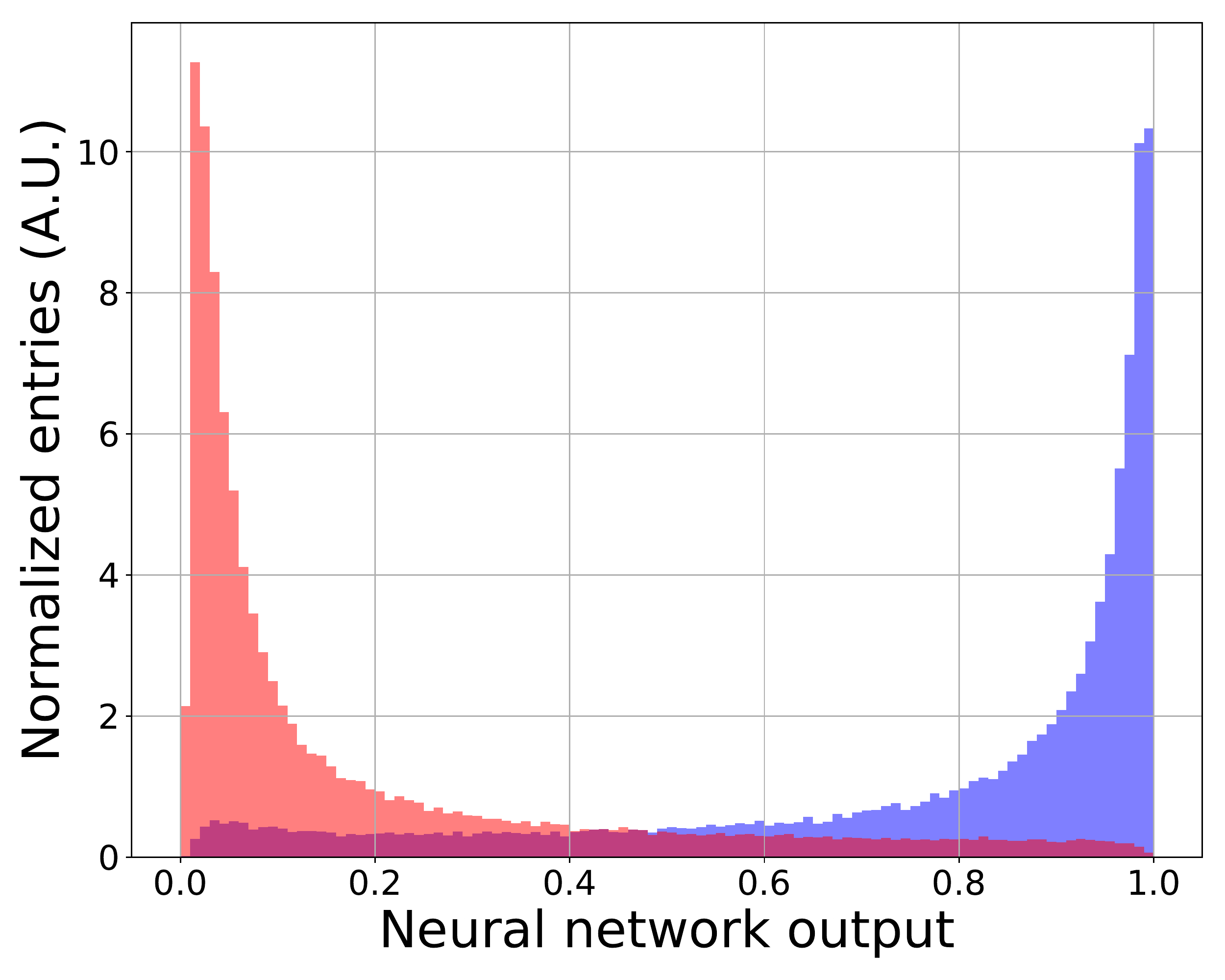}
	\end{center}
	\caption{\label{fig:cnnperf} Left panel: Performance of CNN, compared with other variables for boosted top-jet tagging: Template Tagger~\cite{Almeida:2010pa,Almeida:2011aa,Backovic:2012jk} and $N$-subjettiness~\cite{Thaler:2010tr,Thaler:2011gf}. Right panel: Neural network output distributions for the signal and background.}
\end{figure}

To train the network, we use the MC samples of particle-level top and QCD jets that were previously used in Ref.~\cite{Almeida:2015jua}. The samples were generated using {\tt MadGraph 5}~\cite{Alwall:2011uj} at parton level ($t\bar{t}$ and $q\bar{q}$ final states), following by showering and hadronization with {\tt Pythia 8}~\cite{Sjostrand:2007gs}. We use a sample of jets reconstructed using anti-$k_T$ algorithm with a large jet cone of $R=1.0$, with $p_T$ in $800 - 900\ {\rm GeV}$ range. We further require that the jet mass ($m_J$) be in the range of $130-210\ {\rm GeV}$. A majority of top-jets would fall in this mass range, while most QCD jets would be rejected by the $m_J$ requirement. The top and QCD jets passing these basic cuts are preprocessed as described in~\cite{Almeida:2015jua}, and provided as inputs to the CNN. For training, weights were initialized using Xavier initialization~\cite{Xavier}, and Adam hyperparameter optimization~\cite{DBLP:journals/corr/KingmaB14} with learning rate of 0.05 and dropout regularization rate of 0.0001 was used. 100 epochs through the training samples were made using minibatches of 1000 events. After training, the CNN performance was evaluated using the test samples, generated and preprocessed in the same way as the training samples.  The results are shown in Fig.~\ref{fig:cnnperf}. The results are not very sensitive on the choices of hyperparameters during training. The CNN output provides a clear separation between the two types of jets. For interesting top tagging efficiencies, the mistag rate is reduced by a factor of almost 2 compared to traditional observables, providing further improvement upon the two hidden layer multi-layer perceptron type deep neural network used in~\cite{Almeida:2015jua}. 

As explained in the Introduction, the notion of IRC safety applies to observables defined on parton-level events. The observable we want to study is the output of the CNN, the ``topness" $Y\in [0, 1]$. The CNN maps a set of energy deposits in individual pixels into this observable: $I(i, j) \to Y$. A parton-level event of the type shown in Fig.~\ref{fig:TopGem} is trivially mapped into $I(i, j)$ by identifying each parton's location in the $(\eta, \phi)$ space, and assigning the value of that parton's energy to the corresponding HCAL cell. This defines the action of the CNN on parton-level events, which can be thought of as a map
\beq
{\cal O}: \{ p_i, i=1\ldots N_p\} \to Y, 
\eeq{CNNpartonmap}
where $p_i$ is the parton 4-momenta, and $N_p$ is the number of partons in the event. We would like to study whether the IRC safety criterion, Eq.~\leqn{IRS_def}, applies to this map. While ${\cal O}$ is a completely well-defined function, it is horrendously complicated and highly non-linear, making an analytic study of its limits impractical. Instead, we will check the IRC safety criterion numerically. To this end, we used {\tt MadGraph}~\cite{Alwall:2011uj} to generate a parton-level sample of hadronically decaying top quarks, with an additional gluon in the final state, as in Fig.~\ref{fig:TopGem} (b). (To avoid unnecessary complexity, we simulate a process with no other colored particles in the final state.) In this simulation, cuts on the gluon momentum and its separation from each quark must be imposed to avoid infinities associated with soft and collinear singularities. Since we are primarily interested in precisely the gluons in the soft and collinear regions, the cuts we impose are very low: $p_T\geq 5$~GeV, $\Delta R_{qg}\geq 0.05$. One may question whether a fixed-order simulation correctly approximates the cross section for such low values of $p_T$ and $\Delta R_{qg}$. For our purposes, however, this question is irrelevant. We want to study how the CNN response is affected by the presence of a soft or collinear gluon, and the purpose of the simulation is simply to provide a sample of such events; we do not use any information about their overall cross section or phase-space distribution.   

To ensure that the CNN is applied in the same regime where it was trained, we compute the ``jet $p_T$" (the sum of the four parton $p_T$'s) and the ``jet invariant mass" (the total invariant mass of the four partons) for each event, and apply the same cuts as in the training sample, $p_T \in [800 - 900]$~GeV, $m_J\in [130-210]$~GeV. The sample constructed in this way is referred to as the {\it unmerged sample}. We construct the {\it merged sample} by taking each event in the unmerged sample, identifying the quark closest to the gluon (in terms of $\Delta R_{ij}$ separation), and replacing that quark and the gluon with a single parton with 4-momentum equal to the sum of the two. Applying the CNN map to the unmerged and merged samples corresponds to evaluating the left-hand side and the right-hand side of Eq.~\leqn{IRS_def}, respectively. Checking the IRC safety criterion then amounts to comparing the CNN outputs on these two samples.

\section{Results}
\label{sec:results}
\begin{figure}[t!]
	\begin{center}
		\includegraphics[width=8cm]{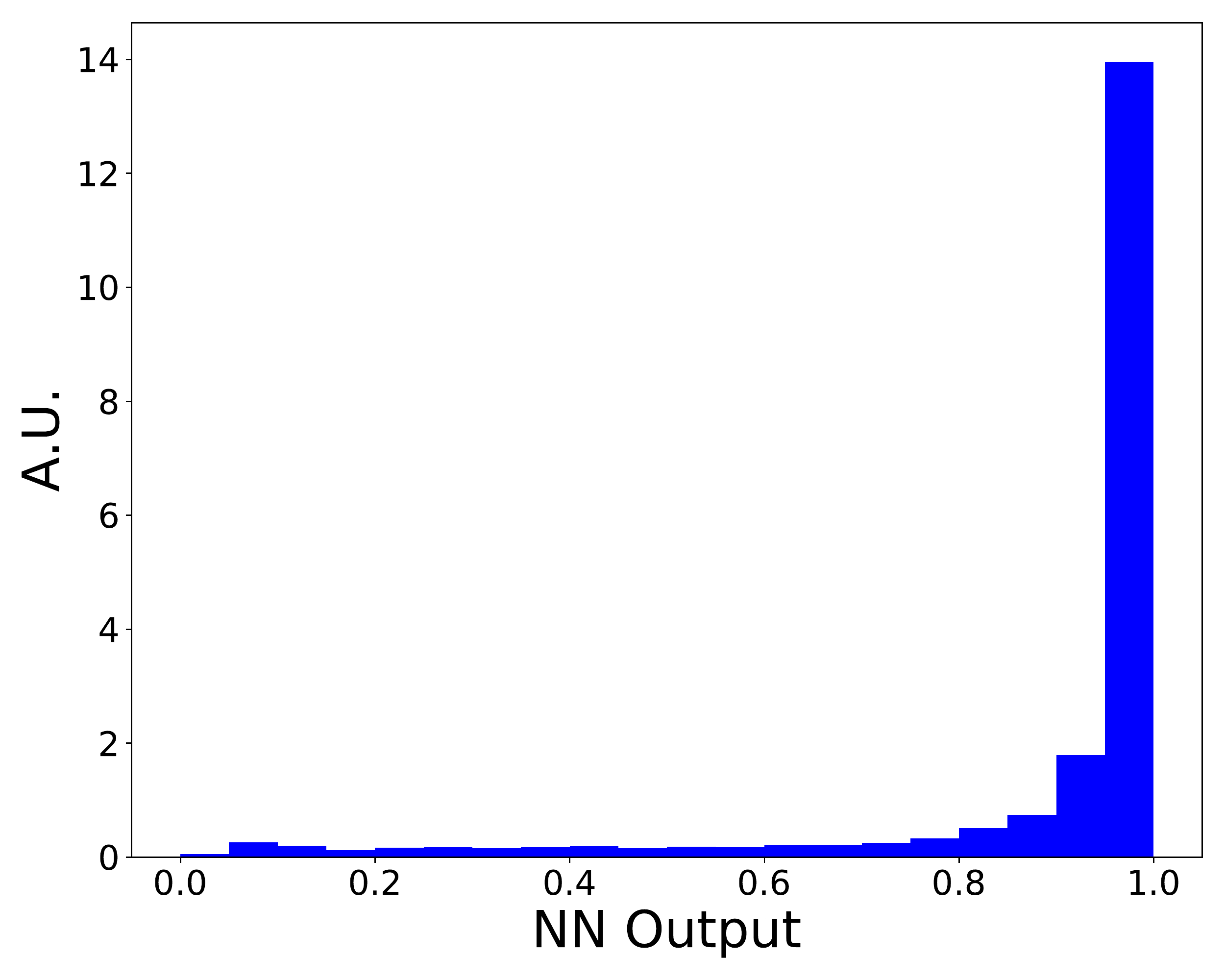}
	\end{center}
\vskip-0.5cm
	\caption{\label{fig:CNNonparton} Neural network output distribution on parton-level top sample.}
\end{figure}

Training the CNN on particle-level top and QCD samples and applying it to the parton-level top sample produces the output distribution shown in Fig.~\ref{fig:CNNonparton}. Clearly, the network predominantly still perceives such events as top-like. The fact that the NN did not need to be retrained when a switch from particle-level to parton-level input was made already provides some evidence that the observable defined by the CNN is likely IRC-safe. In the rest of this section, we will attempt to establish the IRC safety more directly, by comparing CNN outputs on merged and unmerged samples as explained above. 

\begin{figure}[t!]
	\begin{center}
		\includegraphics[width=0.48\linewidth]{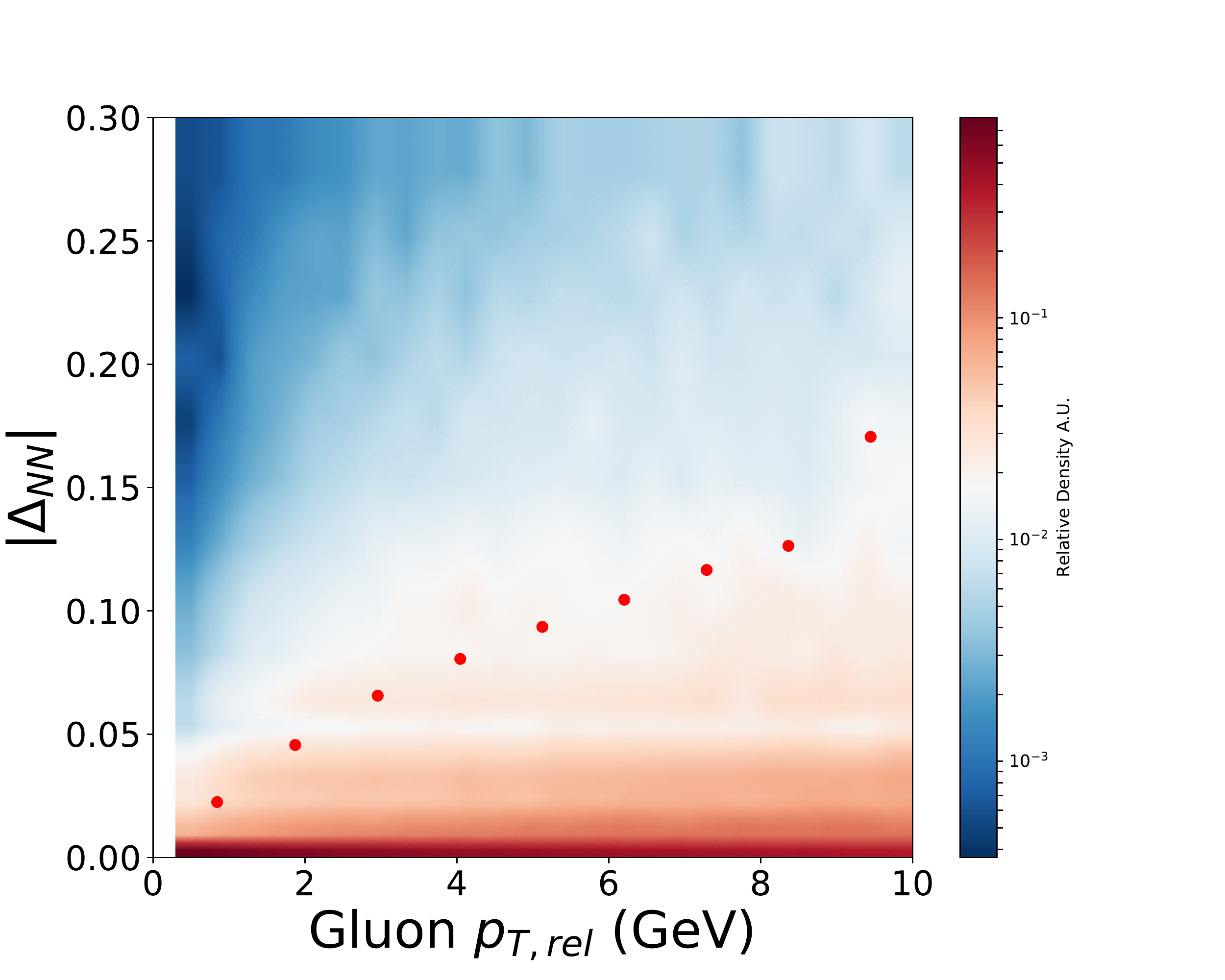}
		\includegraphics[width=0.41\linewidth]{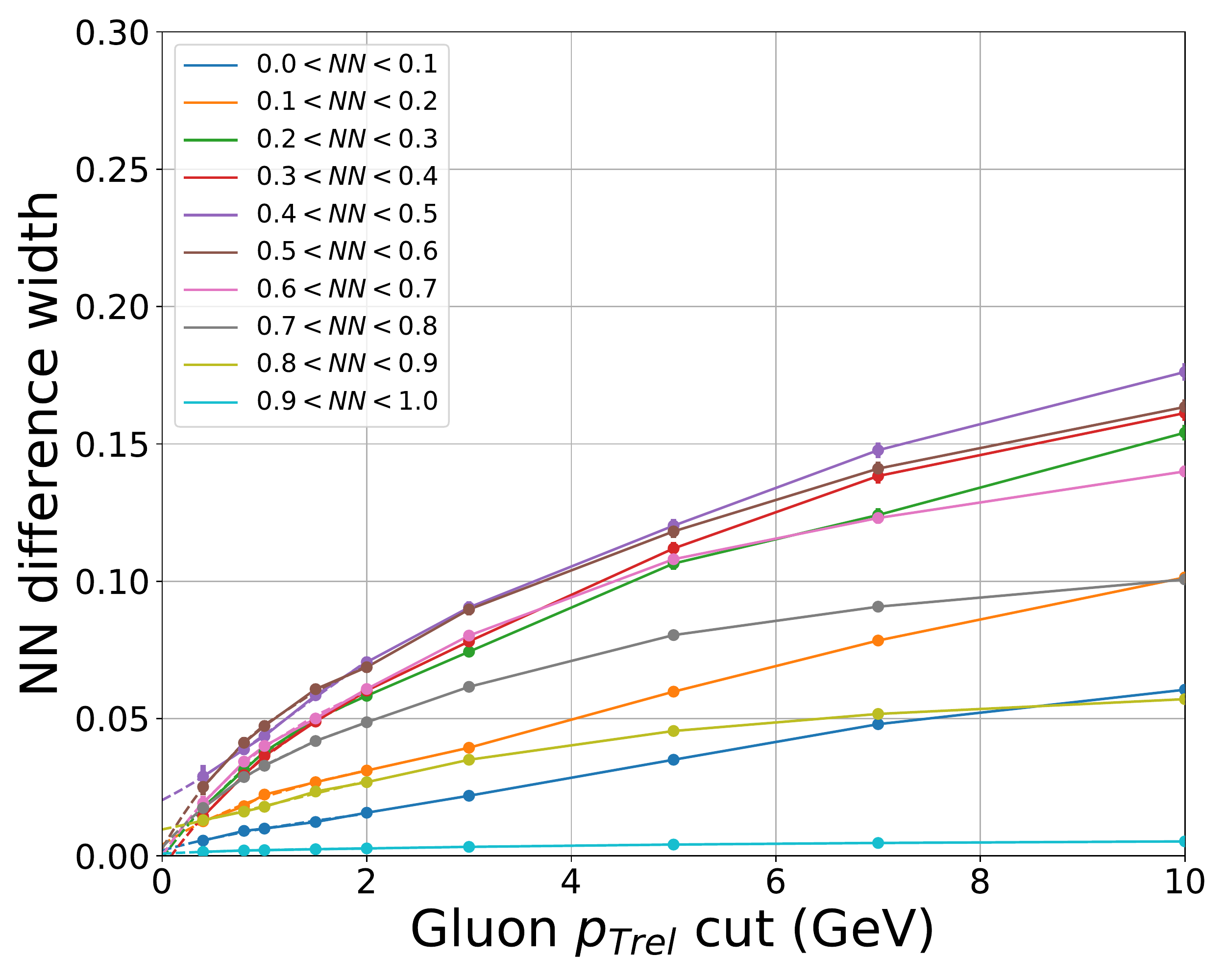}
	\end{center}
	\caption{\label{fig:pTrel} Left panel: Difference in CNN output between merged and unmerged events, $|\Delta_{NN}|$, as a function of the gluon transverse momentum relative to its nearest quark, $p^g_T$. Red dots show the width of the $|\Delta_{NN}|$ distribution. Background colors indicate the relative density of events for given $p^g_T$. Right panel: $|\Delta_{NN}|$ width as a function of $p^g_T$, shown separately for 10 NN output bins. The lines indicate an interpolating curve (third-order polynomial) fit to the data in each NN output bin.}
\end{figure}

To gauge the impact of soft/collinear gluon radiation, we compute the difference $\Delta_{NN}$ between the CNN output from an event in the unmerged sample and the corresponding event in the merged sample. A convenient measure of soft/collinear kinematics of the gluon is provided by its ``relative $p_T$", defined by
\beq
p^g_T = \left|{\bf p}_g-\frac{{\bf p}_g\cdot {\bf p}_q}{|{\bf p}_q|^2}{\bf p}_q\right|,
\eeq{pTg_def}   
where ${\bf p}_q$ is the 3-momentum of the quark nearest (in terms of $\Delta R_{qg}$ separation) to the gluon. Physically, $p^g_T$ is the component of the gluon 3-momentum transverse to the nearest quark, and it vanishes in both soft and collinear limits. If the CNN observable is IRC safe, we expect $\Delta_{NN}$ to go to zero in the limit of vanishing relative $p_T$. The distribution of $|\Delta_{NN}|$ and $p^g_T$ values in our event sample is shown in the left panel of Fig.~\ref{fig:pTrel}. For most events, $|\Delta_{NN}|$ is small, which is reassuring: adding a soft gluon does not lead to a dramatic change in the CNN output. There is, however, a tail of events where the change is significant. To better characterize this tail, we bin the data in relative $p_T$ and calculate the {\it width} of the $|\Delta_{NN}|$ distribution in each bin. The width $|\Delta_{NN}|_{90}$ for each bin is defined by requiring that 90\% of the events in that bin have $|\Delta_{NN}|\leq |\Delta_{NN}|_{90}$. The values of $|\Delta_{NN}|_{90}$ are plotted as red dots in Fig.~\ref{fig:pTrel}. The data exhibits a clear correlation between decreasing relative $p_T$ and decreasing width, indicative of IRC safety. In fact, the data is consistent with the hypothesis that $|\Delta_{NN}|_{90}\to 0$ in the limit of $p^g_T\to 0$. 

In the right panel of Fig.~\ref{fig:pTrel}, the data is further subdivided into 10 bins according to the NN output evaluated on the merged sample, and dependence of the width on relative $p_T$ is shown separately for each bin.\footnote{In some bins, the distribution of $\Delta_{NN}$ is sharply asymmetric around zero, mainly because $Y$ is restricted to lie between 0 and 1 by construction. To account for this, in the right panel of Fig.~\ref{fig:pTrel} we do not take the absolute value of $\Delta_{NN}$, but instead define the ``NN difference width" as the width of the mimimal interval containing 90\% of events, not necessarily centered at 0.}    	
	 For events in the last bin, $0.9\leq Y\leq 1$, emission of an extra gluon has almost no effect even if it has a relatively large relative $p_T$. This is presumably due to the fact that $Y$ is already close to the upper boundary. The events in this bin are therefore consistent with the IRC safety hypothesis, but do not show much variation as relative $p_T$ is varied. On the other hand, events in all other $Y$ bins show a very clear convergence between the output values with and without the extra gluon in the $p^g_T\to 0$ limit.         

\begin{figure}[t!]
	\begin{center}
		\includegraphics[width=0.48 \linewidth]{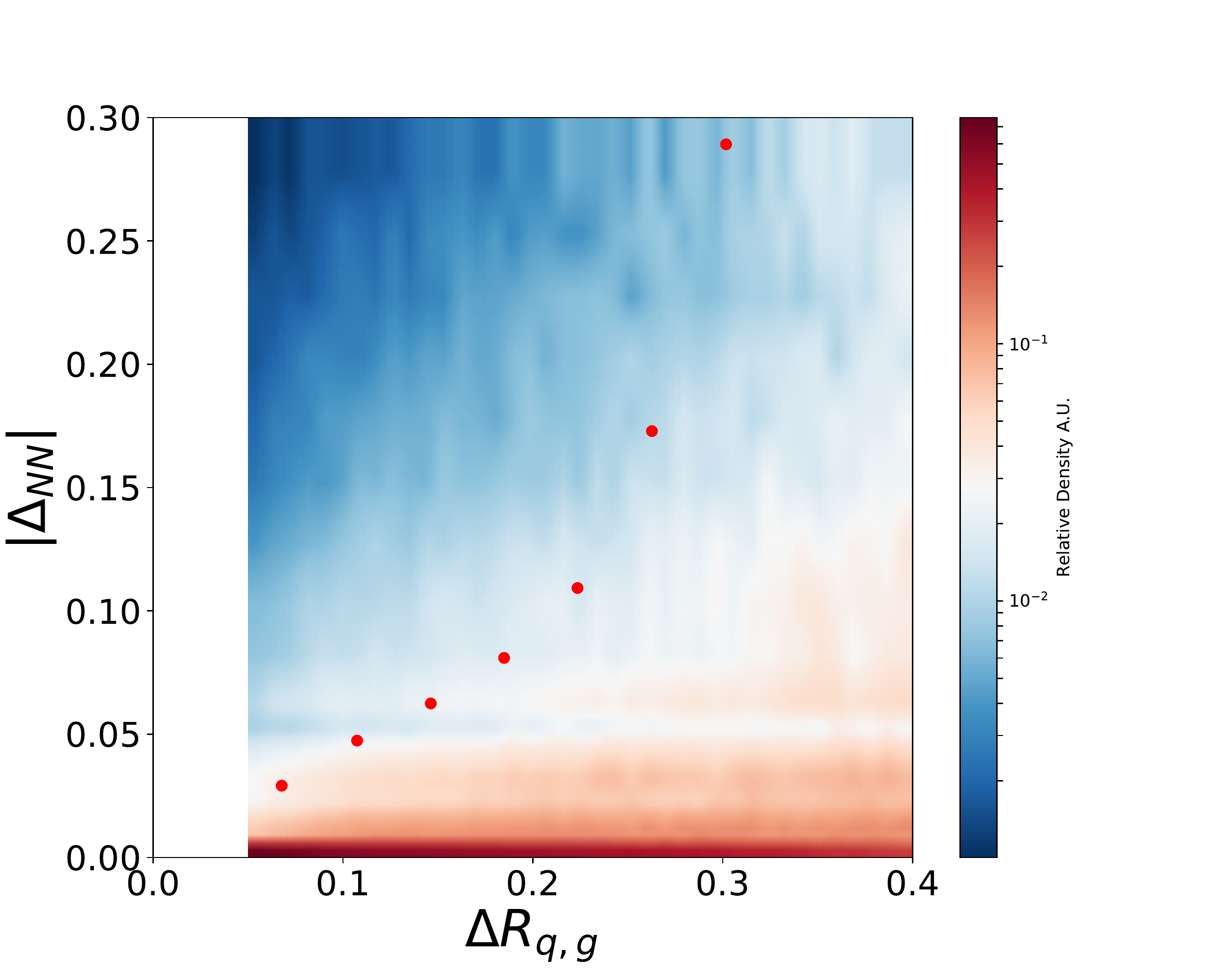}
		\includegraphics[width=0.41 \linewidth]{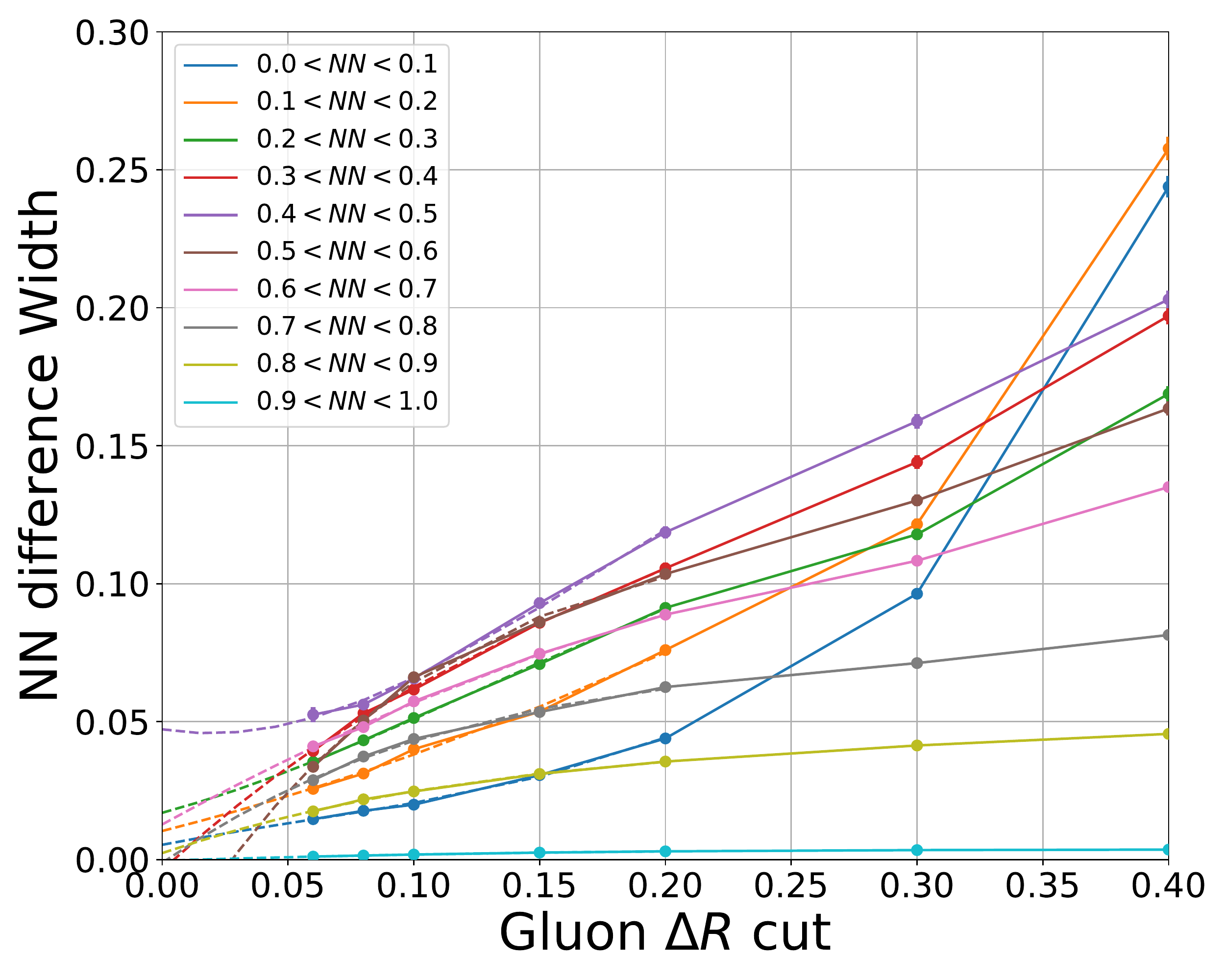}
	\end{center}
	\caption{\label{fig:dR} Left panel: Difference in CNN output between merged and unmerged events, $|\Delta_{NN}|$, as a function of the gluon's angular separation from its nearest quark, $\Delta R_{qg}$. Red dots show the width of the $|\Delta_{NN}|$ distribution. Background colors indicate the relative density of events for given $\Delta R_{qg}$. Right panel: $|\Delta_{NN}|$ width as a function of $\Delta R_{qg}$, binned in 10 NN output intervals. The lines indicate an interpolating curve (third-order polynomial) fit to the data in each NN output bin.}
\end{figure}

\begin{figure}[t!]
	\begin{center}
		\includegraphics[width=0.48 \linewidth]{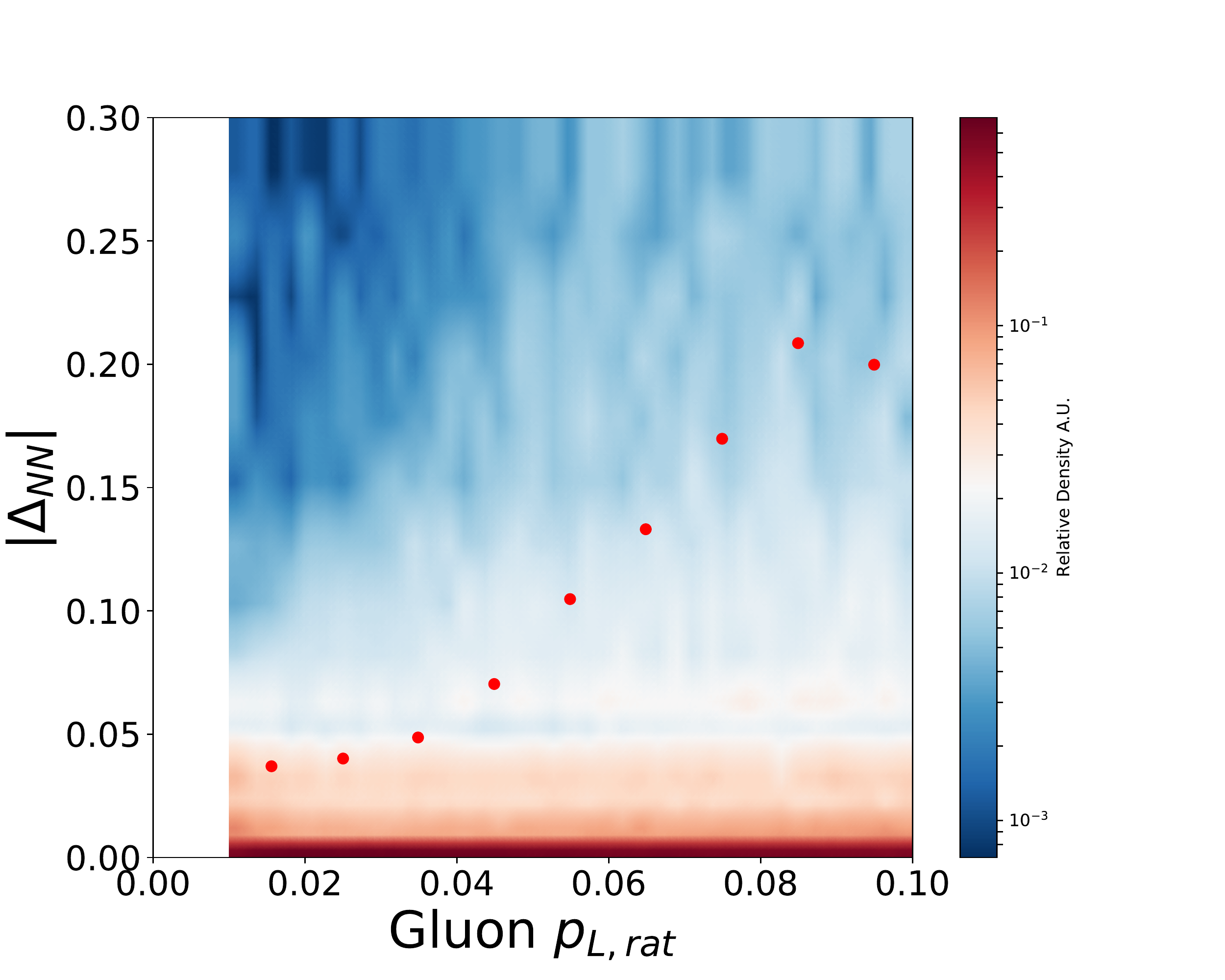}
		\includegraphics[width=0.41 \linewidth]{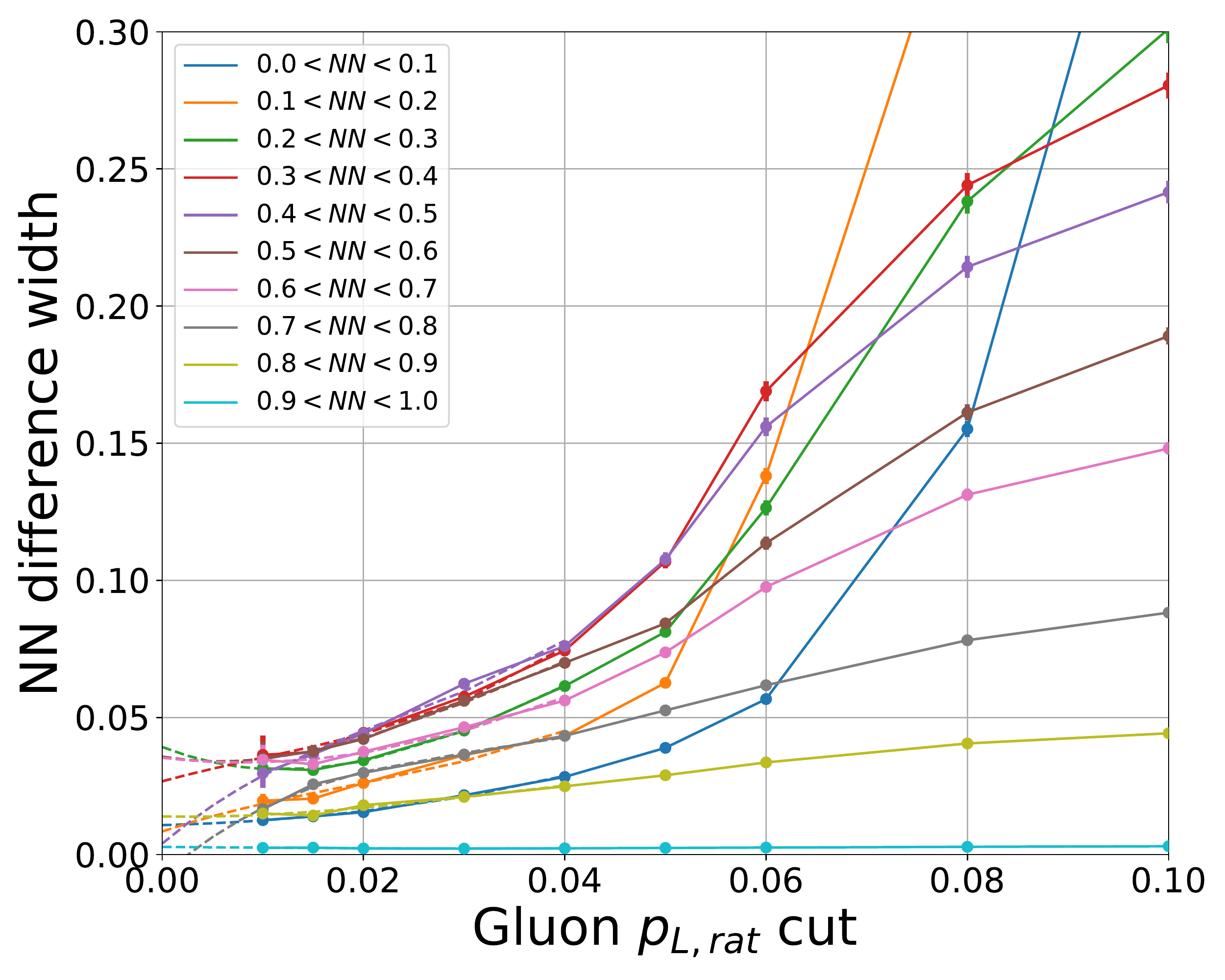}
	\end{center}
	\caption{\label{fig:pL} Left panel: Difference in CNN output between merged and unmerged events, $|\Delta_{NN}|$, as a function of the longitudinal momentum ratio defined in Eq.~\leqn{pLrat}. Red dots show the width of the $|\Delta_{NN}|$ distribution. Background colors indicate the relative density of events for given longitudinal momentum ratio. Right panel: $|\Delta_{NN}|$ width as a function of longitudinal momentum ratio, binned in 10 NN output intervals. The lines indicate an interpolating curve (third-order polynomial) fit to the data in each NN output bin.}
\end{figure}

The relative $p_T$ observable goes to zero in both soft and collinear limits. It is interesting to probe the convergence of the CNN output in each of these limits separately. To this end, we study two observables. The first one is the angular separation between the gluon and the nearest quark, $\Delta R_{qg}$, which goes to zero in the collinear limit, but not the soft limit. The second one is the ``longitudinal momentum ratio", defined by
\beq
p_{L, {\rm rat}}=\frac{{\bf p}_g \cdot {\bf p}_q}{{\bf p}_q \cdot {\bf p}_q},
\eeq{pLrat}
where ${\bf p}_q$ is the 3-momentum of the quark nearest (in terms of $\Delta R_{qg}$ separation) to the gluon. This observable vanishes when the gluon is soft, but not when it is collinear with one of the quarks. The difference in CNN outputs for merged and unmerged samples as a function of these two observables is shown in Figs.~\ref{fig:dR} and~\ref{fig:pL}. We conclude that the convergence of the outputs holds separately in both soft and collinear limits.      

\begin{figure}[t]
	\begin{center}
		\includegraphics[width=0.8 \linewidth]{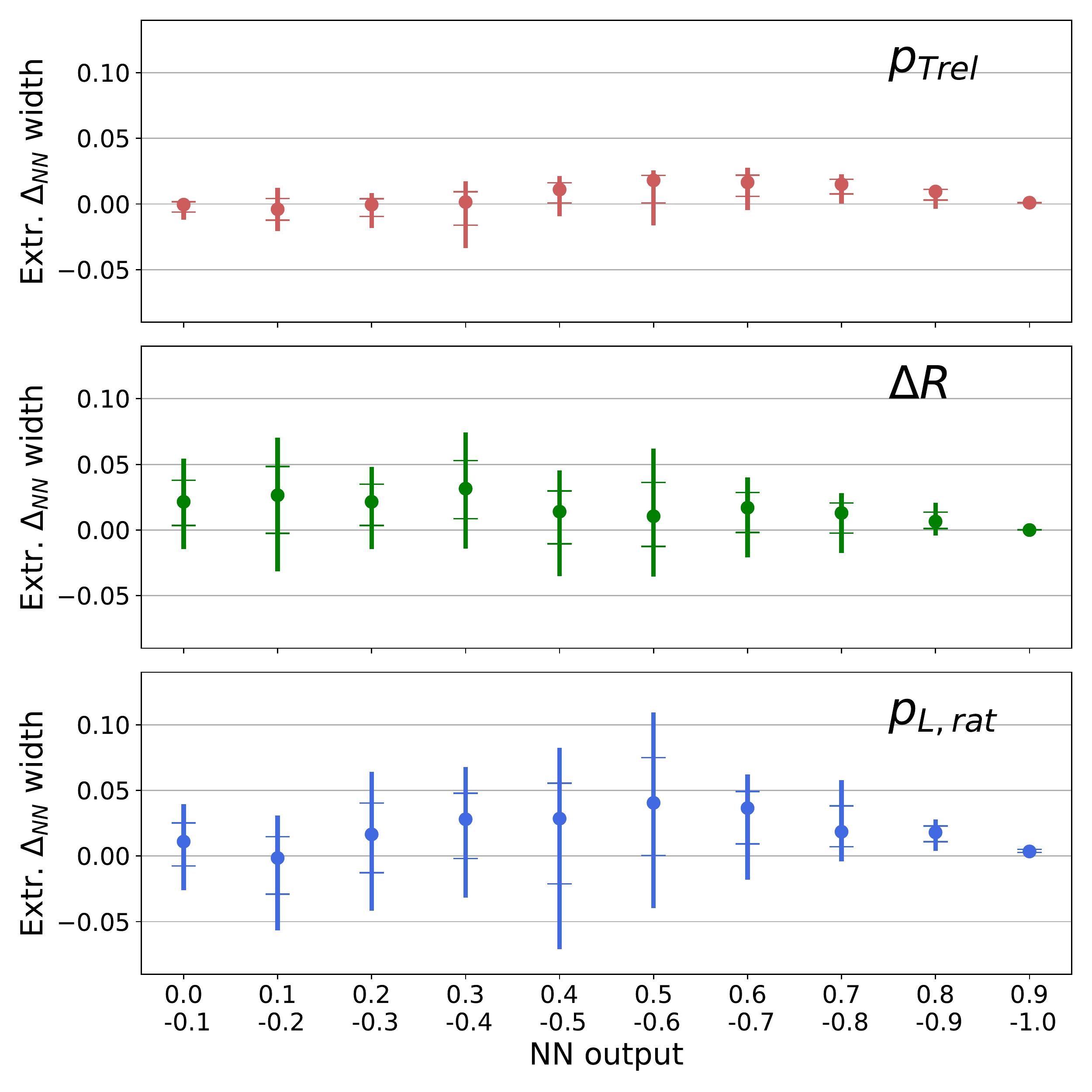}
	\end{center}
	\caption{\label{fig:uncertain} Extrapolated values of $\Delta(0)$, including 68\% and 95\% confidence intervals. Training and extrapolation uncertainties are combined in quadruture.}
\end{figure}

The right panels of Figs.~\ref{fig:pTrel},~\ref{fig:dR}, and~\ref{fig:pL} include interpolating functions obtained by fitting the data in each NN output bin with a third-order polynomial. Expected values of $|\Delta_{NN}|$ in the exact IR(C) limit can be obtained by extrapolating these functions. Let us define $\Delta(x)$ to denote the width of the $|\Delta_{NN}|$ distribution in the bin centered at $x$, where $x=p^g_T$, $\Delta R_{qg}$, or $p_{L, {\rm rat}}$. In all three cases, $x=0$ corresponds to the exact IR(C) limit, so that $\Delta(0)=0$ would indicate exact IR(C) safety. It is clear from Figs.~\ref{fig:pTrel}-\ref{fig:pL} that $\Delta(0)\leq 0.02$ in most cases, and $<0.05$ in all cases. Further, since these values are obtained numerically, they are inherently uncertain. An important uncertainty is introduced by the NN tagger training process, which converges to slightly different weight configurations depending on the random weight initialization at the start of training. To estimate this uncertainty, we trained an ensemble of 20 identical NN taggers, with the only difference being the random initial weights. We then repeated the above analysis for each NN in the ensemble, and used the spread in the values of $\Delta(0)$ measured within this ensemble as an estimate of the training uncertainty. Further, we estimated the uncertainty due to extrapolation from the measured values of $\Delta$ at finite $x$ to $x=0$, and combined it with the training uncertainty in quadrutures to estimate the total uncertainty. The measured values of $\Delta(0)$ including this uncertainty are shown in Fig.~\ref{fig:uncertain}. Our data is consistent with the hypothesis that $\Delta(0)=0$. (The systematic preference for positive offset is due to the definition of $\Delta$, which is defined to be non-negative at all points where it is measured; $\Delta(0)<0$ occurs only as an artifact of extrapolation.)       

Our analysis shows a clear trend towards convergence of the NN output with and without an extra gluon, in the limit when the extra gluon is soft or colinear, and extrapolation to the exact soft/collinear limit is consistent with exact IR(C) safety. The analysis leaves the room open for ${\cal O}$(few~\%) deviations from exact IR(C) safety, due to limitations inherent in the numerical approach to this issue. However, we note that in practice, such small deviations will not change the outcome of the tagging procedure for a vast majority of jets. For example, if the tagging threshold is chosen to be $Y=0.55$, we find that only 0.7\% of events in the top sample lie within $\Delta Y = \pm 0.02$ of the threshold, while 1.8\% lie within $\Delta Y = \pm 0.05$, meaning that the classification of $98-99$\% of top jets would be unaffected.       

\begin{figure}[t]
	\begin{center}
		\includegraphics[width=0.48 \linewidth]{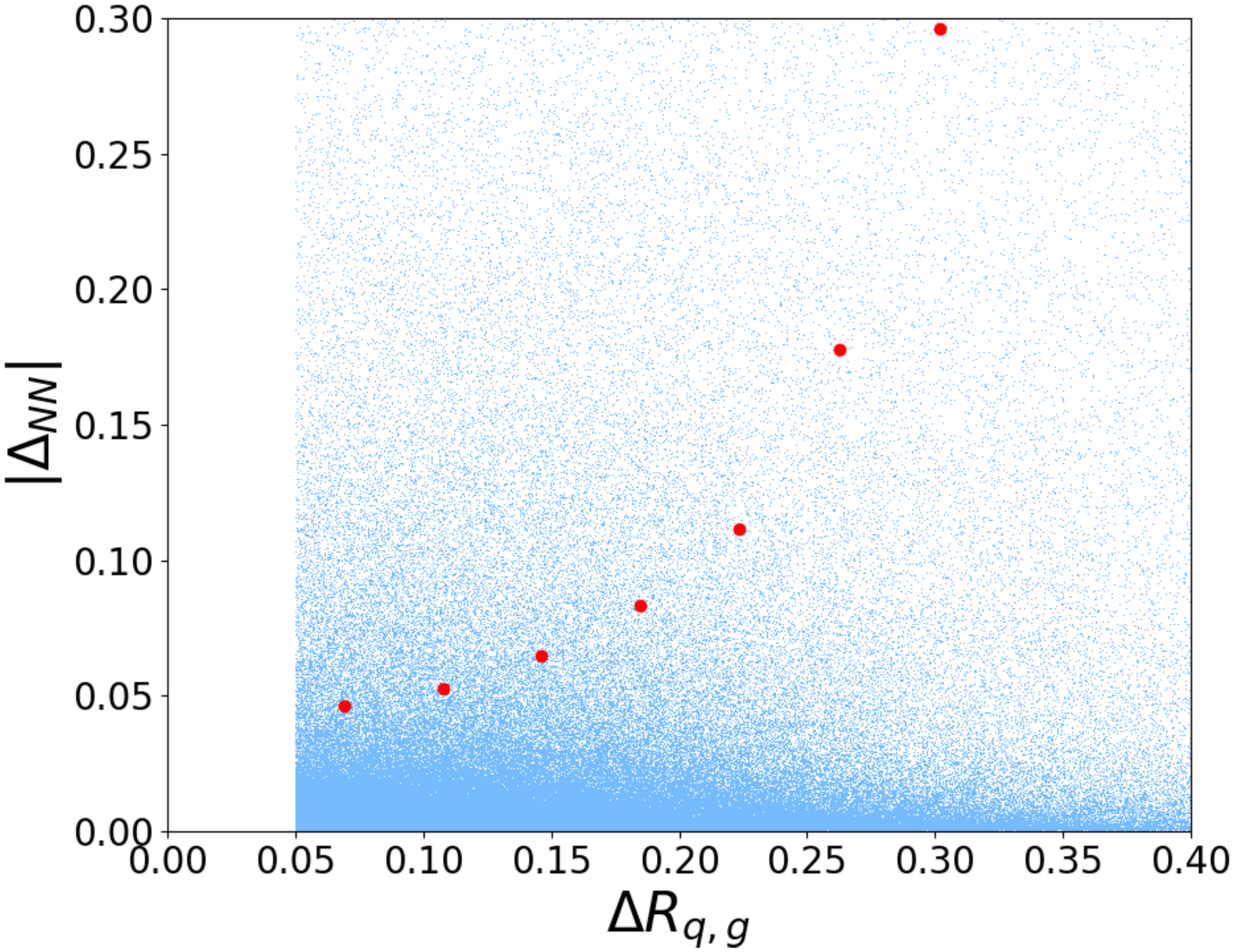}
		\includegraphics[width=0.41 \linewidth]{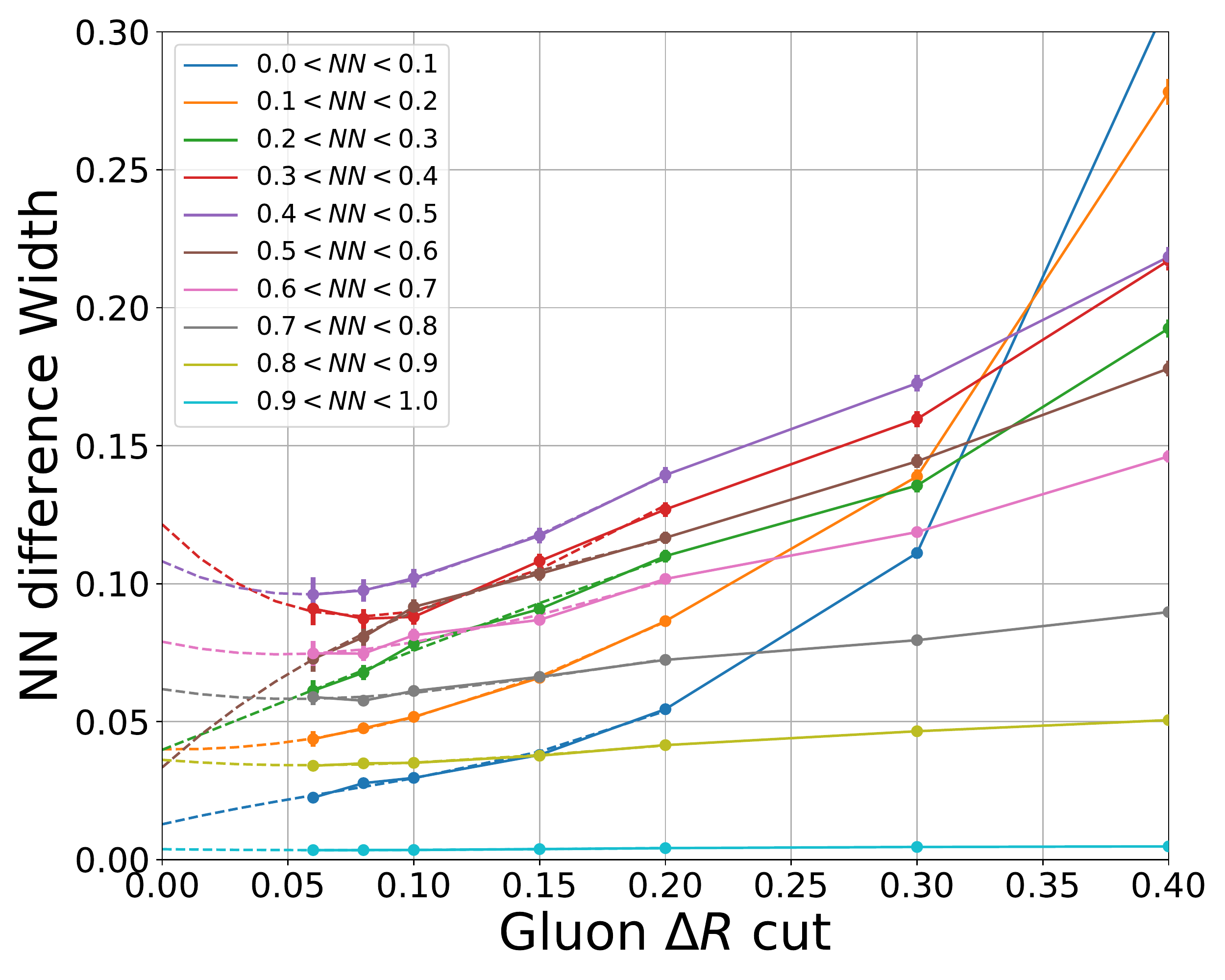}
	\end{center}
	\caption{\label{fig:dR_NOSAMECELL} Same distributions as in Fig.~\ref{fig:dR}. Events where the gluon and its nearest quark are in the same pixel have been removed from the sample.}
\end{figure}

By construction, events in which the extra gluon lands in the same pixel as its nearest quark will have $\Delta_{NN}=0$. This feature makes the CNN observable automatically IRC-safe in the limit of small $\Delta R$. Is the observed IRC safety in this limit due entirely to this feature? To address this question, we repeated the analysis on a sample in which events where the gluon and a quark are in the same pixel have been removed. The result is shown in Fig.~\ref{fig:dR_NOSAMECELL}. The trend towards the CNN outputs converging in the $\Delta R\to 0$ limit persists in this sample. This indicates that the CNN output for a sample with an extra gluon converges smoothly as the gluon approaches its nearest quark, even if they do not land in the same cell. Such convergence is a non-trivial feature of CNN's treatment of energy patterns, and not just a trivial consequence of finite cell size.

\section{Discussion}
\label{sec:discussion}
Starting with Ref.~\cite{Almeida:2015jua}, many studies have demonstrated the efficacy of Neural Networks for boosted top jet tagging, at the level of Monte Carlo (MC) simulations. All studies to date have trained and evaluated NN top taggers using particle-level MC samples of top and QCD jets. In this paper, a Convolutional Neural Network (CNN) top jet tagger was constructed. While particle-level MC samples were used in training as usual, we then applied the resulting tagger to a sample of {\it parton-level} top events with and without an additional gluon in the final state, as shown in Fig.~\ref{fig:TopGem}. We showed that this observable obeys the Infrared Safety criterion: The output of the CNN applied to an event with an extra gluon approaches its output on the same event without an extra gluon, in the limit when the extra gluon becomes soft or collinear with one of the quarks.        

Our analysis does not constitute a complete proof of IRC safety of the CNN output. The reason is that we studied only one class of final states, those containing a high-$p_T$ top quark. In the language of Eq.~\leqn{IRS_def}, our analysis demonstrates IRC safety for some configurations of final-state parton momenta $p_i$, but not for general $p_i$. We did not study the behavior of the CNN output on non-top events, mainly because there is a very broad range of possible momentum configurations, making a comprehensive numerical study impractical. Moreover, it is not immediately clear which of these confugurations are most important for the top tagging problem. We hope to be able to address this issue in future work.

In spite of this limitation, the results of the analysis presented in this paper are highly reassuring. Certainly, the value of NN-based approach to top jet tagging would be in very serious doubt if an addition of a soft or collinear parton to the final state led to an order-one change in the NN output. We showed that this is not the case for the events with a genuine high-$p_T$ top quark in the final state. This result places the NN-based taggers on firmer foundation, and should provide encouragement for further development of this approach.

\section*{Acknowledgements}
The authors are grateful for conversations with Mihailo Backovic, Steven Durr, and Jesse Thaler. MP acknowledges the support of the U.S. National Science Foundation through grant PHY-1719877. SC was supported by National Research Foundation of Korea (NRF) grant NRF-2018R1A2B6005043.
SL was supported by Basic Science Research Program through the National Research Foundation of Korea(NRF) funded by the Ministry of Education (NRF-2018R1D1A1B07049148), by the Korea government (MEST) (NRF-2015R1A2A1A15052408),  and Samsung Science and Technology Foundation under Project Number SSTF-BA1601-07.

%
%

\bibliography{lit}

\providecommand{\href}[2]{#2}\begingroup\raggedright\begin{thebibliography}{10}

\bibitem{Cogan:2014oua}
J.~Cogan, M.~Kagan, E.~Strauss and A.~Schwarztman, \emph{{Jet-Images: Computer
  Vision Inspired Techniques for Jet Tagging}},
  \href{http://dx.doi.org/10.1007/JHEP02(2015)118}{\emph{JHEP} {\bfseries 02}
  (2015) 118}, [\href{https://arxiv.org/abs/1407.5675}{{\ttfamily 1407.5675}}].

\bibitem{Almeida:2015jua}
L.~G. Almeida, M.~Backovic, M.~Cliche, S.~J. Lee and M.~Perelstein,
  \emph{{Playing Tag with ANN: Boosted Top Identification with Pattern
  Recognition}}, \href{http://dx.doi.org/10.1007/JHEP07(2015)086}{\emph{JHEP}
  {\bfseries 07} (2015) 086},
  [\href{https://arxiv.org/abs/1501.05968}{{\ttfamily 1501.05968}}].

\bibitem{deOliveira:2015xxd}
L.~de~Oliveira, M.~Kagan, L.~Mackey, B.~Nachman and A.~Schwartzman,
  \emph{{Jet-images ? deep learning edition}},
  \href{http://dx.doi.org/10.1007/JHEP07(2016)069}{\emph{JHEP} {\bfseries 07}
  (2016) 069}, [\href{https://arxiv.org/abs/1511.05190}{{\ttfamily
  1511.05190}}].

\bibitem{Baldi:2016fql}
P.~Baldi, K.~Bauer, C.~Eng, P.~Sadowski and D.~Whiteson, \emph{{Jet
  Substructure Classification in High-Energy Physics with Deep Neural
  Networks}}, \href{http://dx.doi.org/10.1103/PhysRevD.93.094034}{\emph{Phys.
  Rev.} {\bfseries D93} (2016) 094034},
  [\href{https://arxiv.org/abs/1603.09349}{{\ttfamily 1603.09349}}].

\bibitem{Barnard:2016qma}
J.~Barnard, E.~N. Dawe, M.~J. Dolan and N.~Rajcic, \emph{{Parton Shower
  Uncertainties in Jet Substructure Analyses with Deep Neural Networks}},
  \href{http://dx.doi.org/10.1103/PhysRevD.95.014018}{\emph{Phys. Rev.}
  {\bfseries D95} (2017) 014018},
  [\href{https://arxiv.org/abs/1609.00607}{{\ttfamily 1609.00607}}].

\bibitem{Komiske:2016rsd}
P.~T. Komiske, E.~M. Metodiev and M.~D. Schwartz, \emph{{Deep learning in
  color: towards automated quark/gluon jet discrimination}},
  \href{http://dx.doi.org/10.1007/JHEP01(2017)110}{\emph{JHEP} {\bfseries 01}
  (2017) 110}, [\href{https://arxiv.org/abs/1612.01551}{{\ttfamily
  1612.01551}}].

\bibitem{Kasieczka:2017nvn}
G.~Kasieczka, T.~Plehn, M.~Russell and T.~Schell, \emph{{Deep-learning Top
  Taggers or The End of QCD?}},
  \href{http://dx.doi.org/10.1007/JHEP05(2017)006}{\emph{JHEP} {\bfseries 05}
  (2017) 006}, [\href{https://arxiv.org/abs/1701.08784}{{\ttfamily
  1701.08784}}].

\bibitem{Louppe:2017ipp}
G.~Louppe, K.~Cho, C.~Becot and K.~Cranmer, \emph{{QCD-Aware Recursive Neural
  Networks for Jet Physics}},
  \href{https://arxiv.org/abs/1702.00748}{{\ttfamily 1702.00748}}.

\bibitem{Pearkes:2017hku}
J.~Pearkes, W.~Fedorko, A.~Lister and C.~Gay, \emph{{Jet Constituents for Deep
  Neural Network Based Top Quark Tagging}},
  \href{https://arxiv.org/abs/1704.02124}{{\ttfamily 1704.02124}}.

\bibitem{Butter:2017cot}
A.~Butter, G.~Kasieczka, T.~Plehn and M.~Russell, \emph{{Deep-learned Top
  Tagging with a Lorentz Layer}},
  \href{https://arxiv.org/abs/1707.08966}{{\ttfamily 1707.08966}}.

\bibitem{Metodiev:2017vrx}
E.~M. Metodiev, B.~Nachman and J.~Thaler, \emph{{Classification without labels:
  Learning from mixed samples in high energy physics}},
  \href{http://dx.doi.org/10.1007/JHEP10(2017)174}{\emph{JHEP} {\bfseries 10}
  (2017) 174}, [\href{https://arxiv.org/abs/1708.02949}{{\ttfamily
  1708.02949}}].

\bibitem{Datta:2017lxt}
K.~Datta and A.~J. Larkoski, \emph{{Novel Jet Observables from Machine
  Learning}}, \href{http://dx.doi.org/10.1007/JHEP03(2018)086}{\emph{JHEP}
  {\bfseries 03} (2018) 086},
  [\href{https://arxiv.org/abs/1710.01305}{{\ttfamily 1710.01305}}].

\bibitem{Cheng:2017rdo}
T.~Cheng, \emph{{Recursive Neural Networks in Quark/Gluon Tagging}},
  \href{https://arxiv.org/abs/1711.02633}{{\ttfamily 1711.02633}}.

\bibitem{Egan:2017ojy}
S.~Egan, W.~Fedorko, A.~Lister, J.~Pearkes and C.~Gay, \emph{{Long Short-Term
  Memory (LSTM) networks with jet constituents for boosted top tagging at the
  LHC}},  \href{https://arxiv.org/abs/1711.09059}{{\ttfamily 1711.09059}}.

\bibitem{Luo:2017ncs}
H.~Luo, M.-x. Luo, K.~Wang, T.~Xu and G.~Zhu, \emph{{Quark jet versus gluon
  jet: deep neural networks with high-level features}},
  \href{https://arxiv.org/abs/1712.03634}{{\ttfamily 1712.03634}}.

\bibitem{Macaluso:2018tck}
S.~Macaluso and D.~Shih, \emph{{Pulling Out All the Tops with Computer Vision
  and Deep Learning}},  \href{https://arxiv.org/abs/1803.00107}{{\ttfamily
  1803.00107}}.

\bibitem{Fraser:2018ieu}
K.~Fraser and M.~D. Schwartz, \emph{{Jet Charge and Machine Learning}},
  \href{https://arxiv.org/abs/1803.08066}{{\ttfamily 1803.08066}}.

\bibitem{Roxlo:2018adx}
T.~Roxlo and M.~Reece, \emph{{Opening the black box of neural nets: case
  studies in stop/top discrimination}},
  \href{https://arxiv.org/abs/1804.09278}{{\ttfamily 1804.09278}}.

\bibitem{Collins:2018epr}
J.~H. Collins, K.~Howe and B.~Nachman, \emph{{CWoLa Hunting: Extending the Bump
  Hunt with Machine Learning}},
  \href{https://arxiv.org/abs/1805.02664}{{\ttfamily 1805.02664}}.

\bibitem{Larkoski:2017jix}
A.~J. Larkoski, I.~Moult and B.~Nachman, \emph{{Jet Substructure at the Large
  Hadron Collider: A Review of Recent Advances in Theory and Machine
  Learning}},  \href{https://arxiv.org/abs/1709.04464}{{\ttfamily 1709.04464}}.

\bibitem{Komiske:2018oaa}
P.~T. Komiske, E.~M. Metodiev, B.~Nachman and M.~D. Schwartz, \emph{{Learning
  to Classify from Impure Samples}},
  \href{https://arxiv.org/abs/1801.10158}{{\ttfamily 1801.10158}}.

\bibitem{Larkoski:2013paa}
A.~J. Larkoski and J.~Thaler, \emph{{Unsafe but Calculable: Ratios of
  Angularities in Perturbative QCD}},
  \href{http://dx.doi.org/10.1007/JHEP09(2013)137}{\emph{JHEP} {\bfseries 09}
  (2013) 137}, [\href{https://arxiv.org/abs/1307.1699}{{\ttfamily 1307.1699}}].

\bibitem{Larkoski:2015lea}
A.~J. Larkoski, S.~Marzani and J.~Thaler, \emph{{Sudakov Safety in Perturbative
  QCD}}, \href{http://dx.doi.org/10.1103/PhysRevD.91.111501}{\emph{Phys. Rev.}
  {\bfseries D91} (2015) 111501},
  [\href{https://arxiv.org/abs/1502.01719}{{\ttfamily 1502.01719}}].

\bibitem{mxnet}
``Apache mxnet: A flexible and efficient library for deep learning.''
  \url{http://mxnet.io/}.

\bibitem{Almeida:2010pa}
L.~G. Almeida, S.~J. Lee, G.~Perez, G.~Sterman and I.~Sung, \emph{{Template
  Overlap Method for Massive Jets}},
  \href{http://dx.doi.org/10.1103/PhysRevD.82.054034}{\emph{Phys. Rev.}
  {\bfseries D82} (2010) 054034},
  [\href{https://arxiv.org/abs/1006.2035}{{\ttfamily 1006.2035}}].

\bibitem{Almeida:2011aa}
L.~G. Almeida, O.~Erdogan, J.~Juknevich, S.~J. Lee, G.~Perez and G.~Sterman,
  \emph{{Three-particle templates for a boosted Higgs boson}},
  \href{http://dx.doi.org/10.1103/PhysRevD.85.114046}{\emph{Phys. Rev.}
  {\bfseries D85} (2012) 114046},
  [\href{https://arxiv.org/abs/1112.1957}{{\ttfamily 1112.1957}}].

\bibitem{Backovic:2012jk}
M.~Backović and J.~Juknevich, \emph{{TemplateTagger v1.0.0: A Template
  Matching Tool for Jet Substructure}},
  \href{http://dx.doi.org/10.1016/j.cpc.2013.12.018}{\emph{Comput. Phys.
  Commun.} {\bfseries 185} (2014) 1322--1338},
  [\href{https://arxiv.org/abs/1212.2978}{{\ttfamily 1212.2978}}].

\bibitem{Thaler:2010tr}
J.~Thaler and K.~Van~Tilburg, \emph{{Identifying Boosted Objects with
  N-subjettiness}},
  \href{http://dx.doi.org/10.1007/JHEP03(2011)015}{\emph{JHEP} {\bfseries 03}
  (2011) 015}, [\href{https://arxiv.org/abs/1011.2268}{{\ttfamily 1011.2268}}].

\bibitem{Thaler:2011gf}
J.~Thaler and K.~Van~Tilburg, \emph{{Maximizing Boosted Top Identification by
  Minimizing N-subjettiness}},
  \href{http://dx.doi.org/10.1007/JHEP02(2012)093}{\emph{JHEP} {\bfseries 02}
  (2012) 093}, [\href{https://arxiv.org/abs/1108.2701}{{\ttfamily 1108.2701}}].

\bibitem{Alwall:2011uj}
J.~Alwall, M.~Herquet, F.~Maltoni, O.~Mattelaer and T.~Stelzer, \emph{{MadGraph
  5 : Going Beyond}},
  \href{http://dx.doi.org/10.1007/JHEP06(2011)128}{\emph{JHEP} {\bfseries 06}
  (2011) 128}, [\href{https://arxiv.org/abs/1106.0522}{{\ttfamily 1106.0522}}].

\bibitem{Sjostrand:2007gs}
T.~Sjostrand, S.~Mrenna and P.~Z. Skands, \emph{{A Brief Introduction to PYTHIA
  8.1}}, \href{http://dx.doi.org/10.1016/j.cpc.2008.01.036}{\emph{Comput. Phys.
  Commun.} {\bfseries 178} (2008) 852--867},
  [\href{https://arxiv.org/abs/0710.3820}{{\ttfamily 0710.3820}}].

\bibitem{Xavier}
X.~Glorot and Y.~Bengio, \emph{Understanding the difficulty of training deep
  feedforward neural networks},  2010.

\bibitem{DBLP:journals/corr/KingmaB14}
D.~P. Kingma and J.~Ba, \emph{Adam: {A} method for stochastic optimization},
  {\emph{CoRR} {\bfseries abs/1412.6980} (2014) },
  [\href{https://arxiv.org/abs/1412.6980}{{\ttfamily 1412.6980}}].

\end{thebibliography}\endgroup
\bibliographystyle{jhep}

\end{document}